\documentclass[aps,amssymb,amsmath,twocolumn,secnumarabic]{revtex4}

\usepackage[dvips]{graphicx}
\usepackage{amsfonts}
\usepackage{amsmath}
\usepackage{mathrsfs}
\usepackage[vcentermath]{youngtab}

\usepackage[arrow, matrix, curve]{xy}
\xymatrixcolsep{0.5cm} \xymatrixrowsep{0.5cm}

\def\L2{{\it Fun\/}\bigl(\text{\GL}\bigr)}

\newcommand\whG{\mbox{\textsc{\^g}}}
\newcommand\whH{\mbox{\textsc{\^h}}}
\renewcommand{\H}{\mbox{\textsc{h}}}
\newcommand{\G}{\mbox{\textsc{g}}}

\newcommand{\OSP}[2]{\mbox{\textsc{osp$(#1|#2)$}}}
\newcommand{\PSU}[2]{\mbox{\textsc{psu$(#1|#2)$}}}
\newcommand{\SU}[2]{\mbox{\textsc{su$(#1|#2)$}}}
\newcommand{\U}[2]{\mbox{\textsc{u$(#1|#2)$}}}

\newcommand{\osp}[2]{\ensuremath{\text{osp}(#1|#2)}}
\newcommand{\psu}[2]{\ensuremath{\text{psu}(#1|#2)}}
\newcommand{\su}[2]{\ensuremath{\text{su}(#1|#2)}}

\def\GL{{\rm GL(1$|$1)}}

\newcommand{\g}{\ensuremath{\mathfrak{g}}}
\newcommand{\m}{\ensuremath{\mathfrak{m}}}

\newcommand{\h}{\ensuremath{\mathfrak{h}}}

\newcommand{\f}{\mathfrak{f}}

\def\GL{{\rm GL(1$|$1)}}

\newcommand{\beqa}{\begin{eqnarray}}
\newcommand{\eeqa}{\end{eqnarray}}

\begin{document}

\title{Superspace Parafermions}

\author{Constantin Candu and Volker Schomerus}

\affiliation{
DESY Theory Group, DESY Hamburg, Notkestrasse 85, D-22603 Hamburg,
Germany}

\date{April 2011}

\begin{abstract}
We describe several families of non-unitary coset conformal field
theories that possess truly marginal couplings. These generalize
the known examples of Wess-Zumino-Witten models on supergroups
such as \PSU{n}{n} or \OSP{2n+2}{2n}. Our extension includes
coset space sigma models, affine Toda theories or Gross-Neveu
models which are believed to arise in certain limits.
\end{abstract}

\maketitle

\vspace*{-7cm} \noindent
{\tt DESY 11-058  \hspace*{12.7cm} \phantom{arXiveyymm.nnnn}}\\[4.7cm]

\Yboxdim9pt

\section{Introduction}

Non-unitary 2-dimensional (2d) field theories provide a
rich and important class of models. They describe
 surface statistical mechanics systems
and they underly the construction of exact string
backgrounds. It is very difficult to obtain any
systematic statements about these theories since
many of the established methods heavily rely on
unitarity. Hence very little is known e.g.\ about
the classification of renormalization group fixed
points.

Within the class of non-unitary 2d field
theories, % but also unitary ones~\cite{joerg},
there is a special subset of models that
possess some (hidden) internal supersymmetry, i.e.\
models that admit the action of some (deformed) superalgebra
in field space.
Intuition suggests that supersymmetry makes these
models more accessible. On the other hand, such
supersymmetric 2d field theories still
possess highly relevant applications e.g.\ to the
study of disordered systems~\cite{Efetov:1983xg}
or strings in Anti-deSitter
backgrounds~\cite{Polyakov:2005ss}.

The models we are about to study can be considered as
a vast extension of the $Z_N$ parafermions
discovered by Fateev and Zamolodchikov in
\cite{Fateev:1985mm}. These provide a family of
2d conformal field theories with central
charges $c = 2 - 6/(k+2)$ whose first members are the
well known Ising ($k=2$) and three-states Potts model
($k=3$). A few years later, Fateev argued that $Z_N$
parafermions possess an integrable perturbation by a
relevant field of dimension $h = (k-1)/k$
\cite{Fateev:1990bf}. This perturbation was studied
further in \cite{Fateev:1991bv} and the limit $k\to\infty$
was identified with the the {\scshape o(3)}
sigma model~\cite{ZZ1, ZZ2}. $Z_N$ parafermions were subsequently
shown to arise as special cases of the so-called
GKO coset construction of Goddard, Kent and Olive
\cite{Goddard:1984vk,Goddard:1986ee}. In fact, $Z_N$
parafermions emerge from the GKO coset model
{\scshape su(2)$/$u(1)}. Many of the features that were initially
found for $Z_N$ parafermions could be extended to more
general cosets. The potential relation between perturbed
GKO cosets and conventional coset sigma models has also
been investigated~\cite{Fendley:1999gb,Fendley:2001hc}.

Here we propose to consider several families of
superspace GKO coset models along with their perturbations.
The simplest representatives are the cosets of the
form $\PSU{2}{2}/\SU{1}{2}$. These provide a family
of non-unitary conformal field theories with central
charge $c=-2$ which is independent of the level. In
contrast to their bosonic relatives, the
$\PSU{2}{2}/\SU{1}{2}$
cosets turn out to possess a perturbation of dimension
$h = 1$. We shall show that the corresponding $\beta$-function
vanishes at two loops.

This result can be generalized to other GKO coset
models involving supergroups as follows. Introduce a Lie supergroup $\G$
along with a subgroup $\H$. We shall denote the
corresponding Lie superalgebras by \g\ and \h,
respectively. The linear space $\m = \g / \h$ carries
a representation of $\h$. We assume that both $\g$ and $\h$
are semisimple. Decomposing $\h=\oplus_{\nu=1}^p\h_\nu$ into
simple superalgebras and $\m=\oplus_{\sigma=1}^q\m_\sigma$  into
$\h$-irreducible summands we are able to formulate the
main claim.
\medskip

\noindent
{\it {\bf Proposition}: Suppose the quadratic
Casimir elements $C^{(2)}_\nu$ of $\h_\nu$ on the
representation spaces $\m_\sigma$ vanish, i.e.\
\begin{equation} \label{main}
C^{(2)}_\nu(\m_\sigma) = 0  \ ,
\end{equation}
for all $\nu =1, \dots,p$ and $\sigma=1,\dots,q$. Then
the GKO coset model $\G/\H$ possesses truly marginal
deformations for all values of the level.}
\medskip

It is easy to check that the GKO coset $\PSU{2}{2}/\SU{1}{2}$
satisfies the condition formulated in our
main claim. In this case, $\psu{2}{2}/\su{1}{2}$
is a direct sum of the fundamental
representation of $\su{1}{2}$ and its dual. Both
have vanishing Casimirs. Our result generalizes
previous findings \cite{Quella:2007sg} for Wess-Zumino-Witten
(WZW) models on supergroups with vanishing dual Coxeter number.

In sec.~\ref{sec:cosets} we describe four families
of coset models satisfying the conditions of our main
claim, see cases~(\ref{coset1}--\ref{coset4}). We construct
explicitly the marginal operators. The corresponding $\beta$-functions
are studied in sec.~\ref{sec:beta} where they are shown
to vanish at two loops. Sec.~\ref{sec:final} concludes
with an overview of applications. In particular, we
discuss exactly marginal deformations connecting various
GKO cosets.
Large level limits shed new light on strong-weak coupling
dualities between conformal sigma models and
Toda-models for algebras \osp{2n+2}{2n} and
\psu{n}{n}.

%%%%%%%%%%%%%%%%%%%%%%%%%%%%%%%%%%%%%

\section{Cosets and perturbations}\label{sec:cosets}

%%%%%%%%%%%%%%%%%%%%%%%%%%%%%%%%%%%%%%

We shall denote the WZW model on a supergroup $\G$ by $\whG$.
The GKO construction identifies the commutant of the current
algebra of $\whH$ within
the field space of $\whG$ with the
chiral algebra of a conformal field theory (CFT) known as
the $\G/\H$ GKO coset or, simply, $\whG/\whH$. %the $\whG/\whH$ model.
Given some choice for the level(s) of $\whG$, the
levels $k_\nu$ of the simple factors $\whH_\nu$ in the
denominator are determined through the so-called
embedding index of $\H$ into $\G$, see \cite{dFrancesco}.

Let us show that the $\whG/\whH$ model contains fields of weight
$(h,\bar h) = (1,1)$. These marginal fields are of the
type $(0,\m_\sigma)$ in standard GKO coset %CFT
notations. In order
to construct them we introduce a basis
$\{j_a\}_{a=1}^{\dim\mathfrak{g}}$ of $\mathfrak{g}$
which respects the grading of $\g$ and the decomposition
\begin{equation}\label{eq:dir_sum}
 \mathfrak{g}\big\vert_\mathfrak{h}=\bigoplus_{\nu=1}^p\mathfrak{h}_\nu
 \oplus\bigoplus_{\sigma=1}^q\mathfrak{m}_\sigma
\end{equation}
of $\mathfrak{g}$ into irreducible representations of $\mathfrak{h}$.
This means that every basis vector $j_a$ is either bosonic or fermionic, that is has degree $|a|=0$ or $|a|=1$ respectively, and belongs to some summand in
eq.~\eqref{eq:dir_sum}. For clarity, we shall sometimes attach
superscripts $\G$, $\H$ and $\G/\H$ to the fields of the
$\whG$, $\whH$ and $\whG/\whH$ models.

The marginal fields $(0,\mathfrak{m}_\sigma)$ are obtained by
decomposing the chiral currents $J_a^{\mathrm{G}}$ indexed by $j_a \in \m_\sigma$
into a product of fields of the
$\whH$ model and of the $\whG/\whH$ model,
\begin{equation} \label{decomp1}
J^{\mathrm{G}}_a(z)
\ = \ \psi^{\mathrm{G}/\mathrm{H}}_\sigma(z) \ \otimes
\ \phi^{\mathrm{H}}_{a}(z)\ ,\ j_a\in\mathfrak{m}_\sigma\ .
\end{equation}
Since the operator product expansion (OPE) between currents
$\{J_k(w)\}_{j_k\in\mathfrak{h}}$ and $\{J_a(z)\}
_{j_a\in\mathfrak{m}_\sigma}$ only contains simple poles,
the same holds for the fields $\phi_a$ on the right hand
side. Equivalently, the fields $\{\phi_a\}_{j_a\in \mathfrak{m}_\sigma}$ are primary fields of $\whH$ within  a single affine
conformal block. The parafermion fields $\{\psi_\sigma\}_{\sigma = 1}^q$ are primaries of the $\whG/\whH$ model. Equating the conformal weights on both sides
of the decomposition formula \eqref{decomp1}
we conclude that the fields $\psi_\sigma$ are indeed
marginal
\begin{equation}\label{eq:marg}
    h(0,\m_\sigma) = 1 - {\sum}_\nu
\frac{C^{(2)}_\nu (\m_\sigma)}{k_\nu + c^\vee_\nu} = 1 \ ,
\end{equation}
where $c^\vee_\nu$ is the dual Coxeter number of $\mathfrak{h}_\nu$.
In the second step we have inserted the key assumption \eqref{main}
of our proposition. We can now deform the $\whG/\whH$ model by
introducing a marginal perturbation
\begin{equation}\label{eq:pert}
{\cal O}(z,\bar z)  = {\sum}_\sigma\lambda^\sigma \mathcal{O}_\sigma(z\,\bar z)={\sum}_\sigma \lambda^\sigma \psi_\sigma(z)
\overline{\psi}_{\bar\sigma}(\bar z)\ ,
\end{equation}
with couplings $\lambda_{\sigma}$. The fields $\overline{\psi}_{\bar\sigma}(\bar{z})$
are constructed similarly from the anti-chiral currents
$\bar{J}_a(\bar z)$, but for $j_a\in\mathfrak{m}_\sigma^*$.%
\smallskip%

\noindent
{\bf Examples:} Suppose that $\H$ is the fixed point set of some
involution of $\G$, that is the pair $(\G,\H)$ corresponds to a
symmetric superspace. Then there are four families of coset models
that satisfy the conditions in~\ref{main}, see also \cite{Canduthese}.
The central charge is either $c=1$ or $c=-2$ depending on wether $\G$
is of {\scshape osp} or respectively {\scshape psu} type.
\smallskip

%\begin{enumerate}
1. Coset theories of real super-Grassmannian type
\begin{equation} \label{coset1}
  \frac{\OSP{2n+2 m+2}{2n+2m}_k}{\OSP{2n+1}{2n}_k
\times \OSP{2m+1}{2m}}_k\ .
\end{equation}
In this case, $\m$ is a bi\-fundamental representation
%of the denominator supergroup
%
$$ \m = \ \young(o)\,^{2m+1}_{2m} \times \young(o)\,^{2n+1}_{2n} \ ,  $$
where $\young(o)\,^{m}_{2n}$ denotes the $\osp{m}{2n}$ fundamental
representation with Casimir  $m-2n-1$. Since $\m$ is
irreducible, there is one marginal parafermion $\psi$.
These are superspace generalizations of the
Grassmanian
parafermions~\cite{Fateev:1991bv, Fendley:2001hc, Fendley:1999gb}.
\smallskip

2. Coset theories of complex super-Grassmannian type
\begin{equation} \label{coset2}
\frac{\PSU{n+m}{n+m}_k}{\SU{n-1}{n}_k \times \SU{m+1}{m}_k}\ .
\end{equation}
For such coset models, $\m$ is only \emph{real} irreducible
$$
\m = \ \young(u)\,^{n-1}_{n} \times \overline{\young(u)}\,^{m+1}_{m} \
\oplus\  \overline{
\young(u)}\,^{n-1}_{n} \times \young(u)\,^{m+1}_{m} $$
where $\young(u)\,^{m}_{n}$ denotes the fundamental representation
of $\SU{m}{n} \cong \SU{n}{m}$ and its dual carries an additional
bar. In both representations, the Casimir assumes the value
$m-n-1/(m-n)$. There are two marginal parafermions
$\psi_+$ and $\psi_-$ related by
the conjugation symmetry $\psi_+^\dagger = \psi_-$ inherited from
the WZW models. The perturbation~\eqref{eq:pert} respects this
symmetry, provided we impose $\lambda_+^*=\lambda_-$ on the
complex couplings $\lambda_\pm$. This family is a superspace
generalization of the usual {\scshape su(2)$/$u(1)} parafermions. The potential
relationship between the $k\to\infty$
limit of this type of perturbed coset CFTs
and the complex Grassmaniann
$\sigma$-models with $\theta$-term was suggested
in~\cite{Fateev:1991bv}.\smallskip

3. Our third series consists of coset models of the form
\begin{equation}   \label{coset3}
  \frac{\PSU{2n}{2n}_k}{\OSP{2n}{2n}_{2k}} \ .
\end{equation}
The representation $\mathfrak{m}$ is irreducible for $n>1$
$$ \m = \young(oo)\,^{2n}_{2n}\  $$
and is obtained from the $\osp{2n}{2n}$ traceless symmetric
tensors of rank 2 by factoring out a 1-dimensional submodule.
The reducible but indecomposable traceless tensor of rank 2 has
a vanishing Casimir eigenvalue, because its restriction to the
1-dimensional submodule vanishes. For $n=1$ we get the particular case of  the series~\eqref{coset2} with $m=n=1$ since $\OSP{2}{2}_{2k}\simeq \SU{2}{1}_k$.
\smallskip

4. Diagonal cosets with $\G=\H\times \H$ and $\H$ simple, i.e.
\begin{equation}\label{coset4}
 \frac{\H_k\times \H_l}{\H_{k+l}}\ .
\end{equation}
Since $\m\simeq\h$ in this case, assumption~\ref{main} requires the adjoint representation of $\h$ to have a vanishing Casimir.
This condition constrains the supergroup $\H$ to be $\PSU{n}{n}$, $\OSP{2n+2}{2n}$ or {\scshape d$(2,1;\alpha)$}.
\smallskip

For the cosets~(\ref{coset1}--\ref{coset4}) one can check that the operator algebra of the parafermions $\psi_\sigma$ cannot generate other non-trivial fields of dimension $h\leq 1$, provided the level of $\whG$ is large enough. This makes the perturbation~\eqref{eq:pert} renormalizable.
Notice that $\h$-invariant level one descendents of $\whG$
primaries with vanishing conformal weight also lead to marginal
fields in the $\whG/\whH$ theory. Of particular interest are the
$q$ fields $\chi_\sigma = (\g,0)_\sigma$ in the
cosets~(\ref{coset1}--\ref{coset3}) and the fields
$\chi_1=(\h\times 0,0)$, $\chi_2=( 0\times \h, 0)$ for the
cosets~\eqref{coset4}. These ``dual'' fields are local with
respect to $\psi_\sigma$ and generate conserved
currents~\cite{Ahn:1990gn} in the $\whG/\whH$ model perturbed
by~\eqref{eq:pert}. Local conserved currents of the type $(0,0)$
and of higher spin can also be constructed~\cite{yang}. These
structures should be crucial to argue that the
perturbation~\eqref{eq:pert} is integrable. As in the case of
the fields $\psi_\sigma$ one may show that the
operator algebra of the fields $\{\chi_\sigma\}_{\sigma=1}^q$
of~(\ref{coset1}--\ref{coset3}) and of the fields $\chi_1$ or
$\chi_2$ of~\eqref{coset4} does not generate other non-trivial
fields of dimension $h\leq 1$. This suggests considering the
dual perturbation built out of $\chi$'s instead of $\psi$'s
on the same footing with~\eqref{eq:pert}. We differ the
analysis of such dual perturbations to our forthcoming work.

%%%%%%%%%%%%%%%%%%%%%%%%%%%%%%%%%%%%
%%%%%%%%%%%%%%%%%%%%%%%%%%%%%%%%%%%%

\section{Vanishing of beta function}\label{sec:beta}

%%%%%%%%%%%%%%%%%%%%%%%%%%%%%%%%%%%%
%%%%%%%%%%%%%%%%%%%%%%%%%%%%%%%%%%%%

The 1-loop $\beta$-function $\beta^{(1)}$ of the CFTs~(\ref{coset1}--\ref{coset4})
perturbed by the operator~\eqref{eq:pert} is given by the coefficient ${C_{\rho\sigma}}^{\tau}$ that multiplies the operator
$\mathcal{O}_\tau$ in the operator product expansion (OPE) of the
product $\mathcal{O}_\rho\mathcal{O}_\sigma$,
\begin{equation}
\beta^\tau = (1-h(0,\m_\tau)) \lambda^\tau -
\frac{1}{2}{C_{\rho\sigma}}^{\tau} \lambda^\rho \lambda^\sigma
+ \dots \ ,
\end{equation}
see e.g.~\cite{Zamolodchikov:1986gt}. The first term is zero due
to eq.~\eqref{eq:marg}.

To see that the second term  vanishes as
well, we adopt the BRST construction of the coset theory in which
the coset fields are represented as cohomology classes in the
direct product $\whG\times\whH$ of WZW models~\cite{Karabali:1989dk}. In the product
theory, our marginal parafermions become
\begin{equation}\label{BRST}
\psi_\sigma(z) = {\sum}_{^{j_a\in\m_\sigma}_
{j_b\in\m_\sigma^*}}
J^{\mathrm{G}}_a(z)
\, \otimes \, \phi^{\mathrm{H}}_{b}(z)
g^{ab}
\ .
\end{equation}
Here we used an invariant non-degenerate form $(\phantom{.},\phantom{.})$ of $\g$ to define $g^{ab}$ as the inverse of the matrix $g_{ab}=(j_a,j_b)$.
The OPE of the fields $\psi_\sigma$ involves the structure constants $f_{ab}^c$ of $\g$ defined as usual $[j_a,j_b]=f_{ab}^{\phantom{}c}j_c$. For symmetric spaces $[\m,\m]\subset\h$. Hence,
$f_{ab}^c$ vanish if $j_a,j_b,j_c \in \m$. This proves that $\beta^{(1)}=0$.

The structure of the 4-point function determines whether scale
invariance is broken at 2-loops. According to Cardy~\cite{Cardy},
the 2-loop $\beta$-function $\beta^{(2)}$ vanishes if $\beta^{(1)}=0$
and only the conformal block of the identity contributes to the 4-point function $\langle \mathcal{O}_\kappa
\mathcal{O}_\rho \mathcal{O}_\sigma \mathcal{O}_\tau\rangle$. We
claim that the 4-point functions in the coset models
(\ref{coset1}--\ref{coset4}) satisfy such a criterion. In the case of
unitary CFTs, Cardy's condition implies that $\psi_\sigma$ are abelian currents and exact marginality follows at all loops.  This conclusion does not extend to our non-unitary models. Nevertheless, we demonstrate that
$\beta^{(2)}=0$ by analyzing the 4-point functions of the perturbing
fields.

For simplicity, we restrict our analysis to the supersphere coset
models~\eqref{coset1}. In order to evaluate the 4-point function
of the the perturbing field~\eqref{BRST} we note that
\begin{align}\label{eq:j_corr}
&\langle J_a\vert J_b(1)J_c(z)\vert J_d\rangle^{\mathrm{G}}\! =\! g_{ab}g_{cd}z^{-2}\!
+ \!(-1)^{|b|\!|c|}
g_{ac}g_{bd} + \\[1mm]
&g_{ad}g_{bc}(1-z)^{-2} + f_{ab}^{\phantom{ab}m}f_{mcd}z^{-1} + f_{bc}^{\phantom{bc}m}f_{amd}(1-z)^{-1}\ , \notag
\end{align}
where we have absorbed the levels into $g_{ab}$ and denoted  $f_{abc}=([j_a,j_b],j_c)$. The 4-point function of the
primary fields $\phi^{\mathrm{H}}$ depends on the cross-ratio $z$ through
\begin{align}\notag
 \langle\phi_a\vert \phi_b(1)\phi_c(z)\vert\phi_d\rangle^{\mathrm{H}}&=g_{ab}g_{cd}F_1(z)+ \\[1mm]
& \hspace*{-2.8cm}  (-1)^{|b|\!|c|}g_{ac}g_{bd}F_2(z)+g_{ad}g_{bc}F_3(z)\ . \label{eq:f_corr}
\end{align}
These terms correspond to the three invariants in the tensor product
$\m^{\otimes 4}$. One may determine the (logarithmic) conformal blocks
$F_i(z)$ with the help of the Knizhnik-Zamolodchikov (KZ)
equation~\cite{dFrancesco}.
%~\cite{Knizhnik198483}.
It implies that
\begin{equation}\label{comb}
F_1(z)+F_2(z)+F_3(z)=1\ .
\end{equation}
Contraction of our eqs.~\eqref{eq:j_corr} and ~\eqref{eq:f_corr} along with
eq.~\eqref{main} eliminates the terms involving $f$, i.e.\ the last two terms
in eq.~\eqref{eq:j_corr}. Then using~\eqref{comb} we obtain
\begin{equation*}
\langle\mathcal{O}\vert \mathcal{O}(1,1) \mathcal{O}(z,\bar{z}) \vert
\mathcal{O}\rangle = |1+z^{-2}+(1-z)^{-2}|^2\ .
\end{equation*}
The answer coincides with the 4-point function of the field $J\bar J$,
where $J$ is an abelian {\scshape u(1)} current. Therefore, by Cardy's criterion $\beta^{(2)}=0$.

The calculation of $\langle \mathcal{O}_\kappa \mathcal{O}_\rho \mathcal{O}_\sigma \mathcal{O}_\tau\rangle$ for the other cosets~(\ref{coset1}-\ref{coset4}) uses a key
fact that follows from eq.~\eqref{main}: {\it{There is a unique $\h$-invariant
form $p$ on the space of invariants in $\m^{\otimes 4}$}}. Let $F_J(z)$ and
$F_\phi(z)$ denote the value of $p$ on the 4-point function of currents and
primaries, respectively. With appropriate normalizations we find
$$\langle\psi_\kappa\vert\psi_\rho(1)\psi_\sigma(z)\vert\psi_\tau\rangle =
F_J(z)F_\phi(z)\ . $$
Application of the KZ equation yields $F_\phi(z)=1$. Furthermore, the last
two terms in eq.~\eqref{eq:j_corr} can be shown to belong to the kernel of
$p$. Hence, in calculating the 4-point function, we may treat
$\psi_\sigma$ as an abelian current. A detailed proof will be
presented in the follow-up paper.

%%%%%%%%%%%%%%%%%%%%%%%%%%%%%%%%%%
%%%%%%%%%%%%%%%%%%%%%%%%%%%%%%%%%%%

\section{Applications and Outlook}\label{sec:final}

%%%%%%%%%%%%%%%%%%%%%%%%%%%%%%%%%%
%%%%%%%%%%%%%%%%%%%%%%%%%%%%%%%%%%%

The classification~(\ref{coset1}--\ref{coset4}) of GKO cosets
with exactly marginal couplings has actually appeared first in
a seemingly different context, namely in the analysis of
$\beta$-functions for sigma models on symmetric superspaces
\cite{Canduthese} that extends and corrects earlier claims
in \cite{Babichenko:2006uc}. This is no accident. It has long
been suspected \cite{Fendley:2001hc,Fateev:1991bv} that perturbed
$\G/\H$ GKO coset models tend to the associated sigma model on the
symmetric space $\G/\H$ in the limit of large levels. Our findings
are in perfect agreement with such a correspondence. In fact, we
have shown that the 2-loop $\beta$-function of the perturbed
$\G/\H$ GKO cosets vanishes if and only if the sigma model on the
symmetric superspace $\G/\H$ is conformal. For the sigma models,
the vanishing of the $\beta$-function was established
non-perturbatively. This
suggests that the same should hold for our perturbed GKO coset
models, a fact we shall prove directly in our forthcoming
paper.

Diagonal cosets $\H_k\times \H_l/\H_{k+l}$ provide the most important
family of GKO coset models. In fact, these theories possess many
relations with other interesting 2d systems. If we set $l=1$ and
let $k\to\infty$ we obtain affine Toda theories for the Lie
(super-)algebra $\h$. Setting $l=2$ instead leads to models with
world-sheet supersymmetry. Increasing $l$ further gives  affine
Toda theories coupled to parafermions~\cite{Bernard:1990ti,Ahn:1990gn}.
We can also consider a limit in which both $k,l\to\infty$ while
keeping $k-l$ fixed. Thereby we obtain the principal chiral
model on the {(super-)}group $\H$ with a WZ term at level $k-l$.
The latter was shown to be conformal in \cite{Bershadsky:1999hk,Berkovits:1999im}.
For a review of various level limits in diagonal cosets
see~\cite{revBernard}.

The relations outlined in the last two paragraphs show that
all previously known examples of superspace models with marginal
deformations can be recovered from our results by
taking appropriate limits. Our findings extend earlier studies
and provide a unifying view that suggests interesting new
connections between known models.

There is much room for dualities between the perturbed GKO cosets~(\ref{coset1}--\ref{coset4}), i.e.\ it might be possible
that some of our coset models can be deformed into each other
by the exactly marginal perturbations we have described.
Such deformations should be considered as superspace
counterparts of the known massless renormalization group
flows between bosonic GKO  cosets. Most famous~\cite{aliosha}
are the level reducing flows of diagonal cosets
$$\frac{\H_k\times \H_l}{\H_{k+l}}\quad
\mathop{\stackrel{UV\phantom{xxxx}\text{scale}\phantom{xxxx} IR}{\xrightarrow{\hspace*{2.5cm}}}}_{\psi\phantom{xx}
\text{perturbation}\phantom{xx}\chi_1}
\quad
\frac{\H_{k-l}\times \H_l}{\H_k}$$
for $k>l$. It seems likely that these are replaced by continuous
deformations of CFTs for the diagonal cosets~\eqref{coset4}.
Flows between non-diagonal cosets have been investigated
e.g.\ by Fendley~\cite{Fendley:2001hc}. Following this work,
we conjecture that  our marginal deformations relate the
coset models\\[-5mm]
{\small
\begin{align*}
\frac{\OSP{2n+2}{2n}_{k}}{\OSP{2n+1}{2n}_{k}}  &\leftrightarrow
\frac{\OSP{2n+2}{2n}_{k-1}  \negthinspace\times \OSP{2n+2}{2n}_{1}}{\OSP{2n+2}{2n}_{k}}\\[2mm]
\frac{\PSU{2n}{2n}_k}{\OSP{2n}{2n}_{2k}}&\leftrightarrow \frac{\PSU{2n}{2n}_{k-1} \times \PSU{2n}{2n}_1}{\PSU{2n}{2n}_{k}}\ .
\end{align*}
}
In the limit $k\to\infty$, the perturbed coset models on
the right hand side tend to the $\OSP{2n+2}{2n}$ and
$\PSU{2n}{2n}$ Gross-Neveu models, respectively. Both theories
are known to be conformal. In the same limit, the first model
on the left hand side approaches the conformal sigma model on
the supersphere $\OSP{2n+2}{2n}/\OSP{2n+1}{2n}$, that is a
superspace with bosonic base the sphere $S^{2n+1}$. The
duality between the supersphere sigma model and the
$\OSP{2n+2}{2n}$ Gross-Neveu model was analyzed in much detail
before \cite{Candu:2008vw,Candu:2008yw,Mitev:2008yt}.
In the second relation, the left hand side tends to
the sigma model on the coset space $\PSU{2N}{2N}/\OSP{2N}{2N}$
which we conjecture to be dual to the $\PSU{2N}{2N}$ Gross-Neveu
model. For the special case $\PSU{2}{2}/\OSP{2}{2}\simeq
\mathbb{C}P^{1|2}$, the duality with the $\PSU{2}{2}$
Gross-Neveu model was also conjectured before~\cite{Ludwig}
but no direct evidence was found in~\cite{Candu:2009ep}.
Given our new findings, it seems worthwhile
revisiting the issue.\ \smallskip

\noindent
{\bf Acknowledgements:}
We would like to thank Stefan
Fredenhagen, Vladimir Mitev, Hubert Saleur and Arkady Tseytlin
for very useful discussions and comments. VS is grateful to the
Mathematical Physics Group of King's College, London, for their
warm hospitality and to the London Mathematical Society for
financial support. CC was supported by SFB676.

%%%%%%%%%%%%%%%%%%%%%%%%%%%%%%%%%%%%

%\section*{Remarks}

%%%%%%%%%%%%%%%%%%%%%%%%%%%%%%%%%%%
%
%\begin{itemize}
% \item Curiously, our superspace parafermions are not para anymore.
%Further things to be discussed (maybe)
% relation with models in Pohlmeyer reductions if there is one.
% level rank dualities
% supersymmetric models.
%\end{itemize}
%


\begin{thebibliography}{99}
%\cite{Fateev:1985mm}

%\bibitem{joerg}
%D.~Ridout and J.~Teschner,
%Integrability of a family of quantum field theories related to sigma models
%arXive:1102.5716

%\cite{Efetov:1983xg}
\bibitem{Efetov:1983xg}
  K.~B.~Efetov,
  %``Supersymmetry and theory of disordered metals,''
  Adv.\ Phys.\  {\bf 32} (1983) 53.
  %%CITATION = ADPHA,32,53;%%

%\cite{Polyakov:2005ss}
\bibitem{Polyakov:2005ss}
  A.~M.~Polyakov,
  %``Supermagnets and sigma models,''
  arXiv:hep-th/0512310.
  %%CITATION = HEP-TH/0512310;%%



\bibitem{Fateev:1985mm}
  V.~A.~Fateev and A.~B.~Zamolodchikov,
  %``Parafermionic Currents In The Two-Dimensional Conformal Quantum Field
  %Theory And Selfdual Critical Points In Z(N) Invariant Statistical Systems,''
  Sov.\ Phys.\ JETP {\bf 62} (1985) 215
  [Zh.\ Eksp.\ Teor.\ Fiz.\  {\bf 89} (1985) 380].
  %%CITATION = ZETFA,89,380;%%

\bibitem{Fateev:1990bf}
  V.~A.~Fateev,
  %``Integrable Deformations In Z(N) Symmetrical Models Of Conformal Quantum
  %Field Theory,''
  Int.\ J.\ Mod.\ Phys.\  {\bf A6} (1991) 2109.
  %%CITATION = IMPAE,A6,2109;%%

%\cite{Fateev:1991bv}
\bibitem{Fateev:1991bv}
  V.~A.~Fateev and A.~B.~Zamolodchikov,
  %``Integrable perturbations of Z(N) parafermion models and O(3) sigma model,''
  Phys.\ Lett. {\bf B271} (1991) 91.
  %%CITATION = PHLTA,B271,91;%%

\bibitem{ZZ1}
A.~B.~Zamoldochikov and Al.~B.~Zamolodchikov,
Ann. Phys. {\bf 120} (1979) 253.

\bibitem{ZZ2}
A.~B.~Zamoldochikov and Al.~B.~Zamolodchikov,
%Massless factorized scattering and sigma models with topological terms
Nucl. Phys. {\bf B379} (1992) 602.
%\cite{Fateev:1990bf}

%\cite{Goddard:1986ee}
\bibitem{Goddard:1986ee}
  P.~Goddard, A.~Kent and D.~I.~Olive,
  %``Unitary Representations Of The Virasoro And Supervirasoro Algebras,''
  Commun.\ Math.\ Phys.\  {\bf 103} (1986) 105.
  %%CITATION = CMPHA,103,105;%%

%\cite{Goddard:1984vk}
\bibitem{Goddard:1984vk}
  P.~Goddard, A.~Kent and D.~I.~Olive,
  %``Virasoro Algebras And Coset Space Models,''
  Phys.\ Lett. {\bf B152} (1985) 88.
  %%CITATION = PHLTA,B152,88;%%

%\cite{Fendley:1999gb}
\bibitem{Fendley:1999gb}
  P.~Fendley,
  %``Sigma models as perturbed conformal field theories,''
  Phys.\ Rev.\ Lett.\  {\bf 83} (1999) 4468.
%  [arXiv:hep-th/9906036].
  %%CITATION = PRLTA,83,4468;%%

%\cite{Fendley:2001hc}
\bibitem{Fendley:2001hc}
  P.~Fendley,
  %``Integrable sigma models and perturbed coset models,''
  JHEP {\bf 0105} (2001) 050.
%  [arXiv:hep-th/0101034].
  %%CITATION = JHEPA,0105,050;%%

%\cite{Quella:2007sg}
\bibitem{Quella:2007sg}
  T.~Quella, V.~Schomerus and T.~Creutzig,
  %``Boundary Spectra in Superspace Sigma-Models,''
  JHEP {\bf 0810} (2008) 024
  [arXiv:0712.3549 [hep-th]].
  %%CITATION = JHEPA,0810,024;%%

\bibitem{dFrancesco}
P.~ Di~Francesco, P. Mathieu and D. S\'en\'echal,
\emph{Conformal Field Theory}, Springer 1997.

\bibitem{Canduthese}
  C.~Candu,
%``Discr\'etisation des mod\'eles sigma invariants conformes sur des supersph\'eres et superespaces projectifs,''
  PhD thesis Universit\'e Paris 6 (2008) 65--75
%l\'Ecole Doctorale de Physique de la R\'egion Parisienne,
[tel.archives-ouvertes.fr/tel-00494973] (in french).

%\cite{Ahn:1990gn}
\bibitem{Ahn:1990gn}
  C.~Ahn, D.~Bernard and A.~LeClair,
  %``Fractional supersymmetries in perturbed coset CFTs and integrable soliton
  %theory,''
  Nucl.\ Phys. {\bf B346} (1990) 409.
  %%CITATION = NUPHA,B346,409;%%

\bibitem{yang}
T.~Eguchi and S.~K.~Yang,
%''Deformations of Conformal Field Theories and Soliton Equations``,
Phys. Lett. {\bf B224} (1989) 373.

\bibitem{Zamolodchikov:1986gt}
A.~B.~Zamolodchikov,
%''Irreversibility of the Flux of the Renormalization Group in a 2D Field Theory``,
JETP Lett. {\bf 43} (1986) 730.
%%CITATION = JTPLA,43,730;%%"

%\cite{Karabali:1989dk}
\bibitem{Karabali:1989dk}
   D.~Karabali and H.~J.~Schnitzer,
   %``BRST Quantization of the Gauged WZW Action and Coset Conformal Field
   %Theories,''
   Nucl.\ Phys. {\bf B329} (1990) 649.
   %%CITATION = NUPHA,B329,649;%%

\bibitem{Cardy}
J.~L.~Cardy,
%''Continuously Varying Exponents and the Value of the Central Charge``,
J. Phys. {\bf A20} (1987) L891.

%\cite{Babichenko:2006uc}
\bibitem{Babichenko:2006uc}
  A.~Babichenko,
  %``Conformal invariance and quantum integrability of sigma models on symmetric
  %superspaces,''
  Phys.\ Lett.\  B {\bf 648} (2007) 254.
%  [arXiv:hep-th/0611214].
  %%CITATION = PHLTA,B648,254;%%

\bibitem{Bernard:1990ti}
D.~Bernard and A.~Leclair,
%"{The Fractional supersymmetric Sine-Gordon models}",
Phys. Lett. {\bf B247} (1990) 309.
%     SLACcitation  = "%%CITATION = PHLTA,B247,309;%%"
%\cite{Candu:2010yg}

%\bibitem{Candu:2010yg}
% C.~Candu, T.~Creutzig, V.~Mitev and V.~Schomerus,
%``Cohomological Reduction of Sigma Models,''
%  JHEP {\bf 1005} (2010) 047
%  [arXiv:1001.1344 [hep-th]].
  %%CITATION = JHEPA,1005,047;%%


%\cite{Bershadsky:1999hk}
\bibitem{Bershadsky:1999hk}
  M.~Bershadsky, S.~Zhukov and A.~Vaintrob,
  %``PSL(n$|$n) sigma model as a conformal field theory,''
  Nucl.\ Phys. {\bf B559} (1999) 205
  [arXiv:hep-th/9902180].
  %%CITATION = NUPHA,B559,205;%%

\bibitem{Berkovits:1999im}
N.~Berkovits, C.~Vafa and E.~Witten,
%`Conformal field theory of AdS background with Ramond-                 Ramond flux}",
JHEP {\bf 9903}  (1999) 018.
%%CITATION = HEP-TH/9902098;%%"

\bibitem{revBernard}
D.~Bernard,
%''On symmetries of some massless 2D field theories``,
Phys. Lett. {\bf B279} (1992) 78.

\bibitem{aliosha}
Al.~B.~Zamolodchikov, Nucl. Phys. {\bf B366} (1991) 122.


%\cite{Candu:2008yw}
\bibitem{Candu:2008yw}
  C.~Candu and H.~Saleur,
  %``A lattice approach to the conformal $\OSP{2S+2}{2S}$ supercoset sigma model.
%  Part II: The boundary spectrum,''
  Nucl.\ Phys. {\bf B808} (2009) 487.
%  [arXiv:0801.0444 [hep-th]].
  %%CITATION = NUPHA,B808,487;%%

%\cite{Candu:2008vw}
\bibitem{Candu:2008vw}
  C.~Candu and H.~Saleur,
  %``A lattice approach to the conformal $\OSP{2S+2}{2S}$ supercoset sigma model. Part I: Algebraic structures in the spin chain. The Brauer algebra,''
  Nucl.\ Phys. {\bf B808} (2009) 441.
%  [arXiv:0801.0430 [hep-th]].
  %%CITATION = NUPHA,B808,441;%%

%\cite{Mitev:2008yt}
\bibitem{Mitev:2008yt}
  V.~Mitev, T.~Quella and V.~Schomerus,
  %``Principal Chiral Model on Superspheres,''
  JHEP {\bf 0811} (2008) 086.
%  [arXiv:0809.1046 [hep-th]].
  %%CITATION = JHEPA,0811,086;%%

\bibitem{Ludwig}
This statement was communicated to us by
H.~Saleur and by A.~Ludwig.

%\cite{Candu:2009ep}
\bibitem{Candu:2009ep}
  C.~Candu, V.~Mitev, T.~Quella, H.~Saleur and V.~Schomerus,
  %``The Sigma Model on Complex Projective Superspaces,''
  JHEP {\bf 1002} (2010) 015.
%  [arXiv:0908.0878 [hep-th]].
  %%CITATION = JHEPA,1002,015;%%




















%\bibitem{Z1fl}
%A.~B.~Zamolodchikov, Int. J. Mod. Phys. {\bf A3} (1988) 746

%\bibitem{cl}
%J.~Cardy, A.~Ludwig, Nucl. Phys. {\bf B285} (1987) 687



\end{thebibliography}
\end{document}